\documentclass[aps,twocolumn,floatfix]{revtex4-1}
\usepackage{bm}
\usepackage{dcolumn}
\usepackage{graphicx}
\usepackage{amsmath}
\usepackage{amssymb} 
\usepackage{amsfonts}
\usepackage{float}
\usepackage{xcolor}
\usepackage[capitalise]{cleveref}

\begin{document}
\title{Chirality and correlations in the spontaneous spin-valley polarization of rhombohedral multilayer graphene} 
\author{Yunsu Jang$^{1,2}$}
\author{Youngju Park$^{3}$}
\author{Jeil Jung$^{3,4}$}
\email{jeiljung@uos.ac.kr}
\author{Hongki Min$^{1,2}$}
\email{hmin@snu.ac.kr}
\affiliation{$^{1}$ Department of Physics and Astronomy, Seoul National University, Seoul 08826, Korea}
\affiliation{$^{2}$ Center for Theoretical Physics, Seoul National University, Seoul 08826, Korea}
\affiliation{$^{3}$ Department of Physics, University of Seoul, Seoul 02504, Korea}
\affiliation{$^{4}$ Department of Smart Cities, University of Seoul, Seoul 02504, Korea}

\date{\today}

\begin{abstract}
We investigate the total energies of spontaneous spin-valley polarized states in bi-, tri-, and tetralayer rhombohedral graphene where the long-range Coulomb correlations are accounted for within the random phase approximation.
Our analysis of the phase diagrams for varying carrier doping and perpendicular electric fields shows that 
the exchange interaction between chiral electrons is the main driver of spin-valley polarization, while the presence of Coulomb correlations brings the flavor polarization phase boundaries to carrier densities close to the complete filling of the Mexican hat shape top at the Dirac points.
We find that the tendency towards spontaneous spin-valley polarization is enhanced with the chirality of the bands and therefore with increasing number of layers. 
\end{abstract}

\maketitle

{\em Introduction.}
Multilayer graphene has a unique stacking-dependent chiral structure~\cite{CastroNeto2009,DasSarma2011,Basov2014,HassanRaza2012},
leading to a series of chiral two-dimensional electron systems (C2DESs) 
describable in terms of their sublattice and layer pseudospin degrees of freedom~\cite{min2008a,min2008b}, which presents exciting opportunities for tailoring multilayer graphene’s electronic behavior through control of the stacking arrangement.
%
The electron-electron interactions in multilayer graphene can give rise to a variety of phenomena~\cite{min2008pseudospin,jung2013gapped,martin2010local,weitz2010broken,grushina2015insulating,nam2016interaction,Yoon2017}. In particular, rhombohedral multilayer graphene provides an excellent platform for studying interaction-induced phenomena as evidenced by recent experimental observations and theoretical studies of superconductivity and correlated phases~\cite{Zhou2021a,Zhou2021b,Zhou2022,Seiler2021,Barrera2021,Szabo2022,Lu2022,Geisenhof2022,Ghazaryan2023,Jimeno-Pozo2023,Qin2023,Patri2023,Dong2023,Lin2023,Shavit2023} due to the flattening of its low-energy dispersion with increasing number of layers.
%
%

\begin{figure}[tb]
\includegraphics[width=1\linewidth]{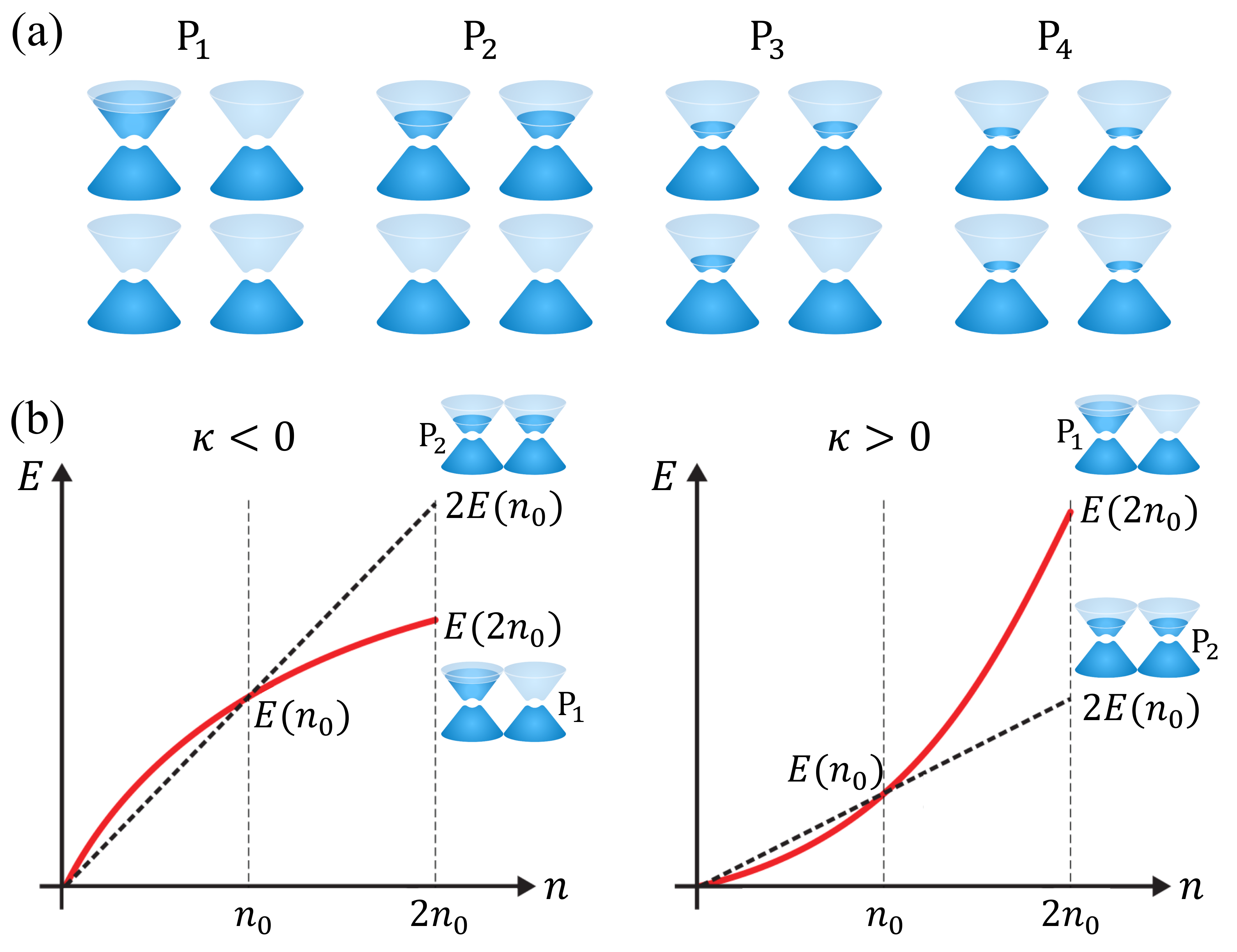}
\caption{
(a) Schematic picture of the spin-valley polarized states ${\rm P}_1 \textendash {\rm P}_4$ that can result from any combination of $K$ and $K'$ valley and $\uparrow$ and $\downarrow$ spin polarizations.
(b) Shape of the density dependence of the energy and the corresponding tendency toward spin-valley polarization. When the density dependence of the energy is concave down (up), 
$E(2n_0)<2E(n_0)$ [$E(2n_0)>2E(n_0)]$ for a density $n_0$ as in the left (right) panel; 
thus the system shows a tendency toward (against) spontaneous spin-valley polarization.
} 
\label{fig:fig1}
\end{figure}

Here, we investigate the spontaneous spin-valley flavor polarization of rhombohedral multilayer graphene for finite carrier doping and perpendicular electric fields by calculating the polarization-dependent total energies including the exact exchange and long-range Coulomb correlations in the random phase approximation (RPA). 
In graphene, there are four distinct spin-valley flavors, ($\uparrow,K$), ($\downarrow,K$), ($\uparrow,K'$), and ($\downarrow,K'$). When electron-electron interactions are absent, these four flavors are occupied equally in the presence of carrier doping. However, electron-electron interactions can break this spin-valley population symmetry, resulting in electrons occupying only one or a subset of the available flavors~\cite{jung2015},
as shown in Fig.~\ref{fig:fig1}.

\begin{figure*}[tb]
\includegraphics[width=0.95\linewidth]{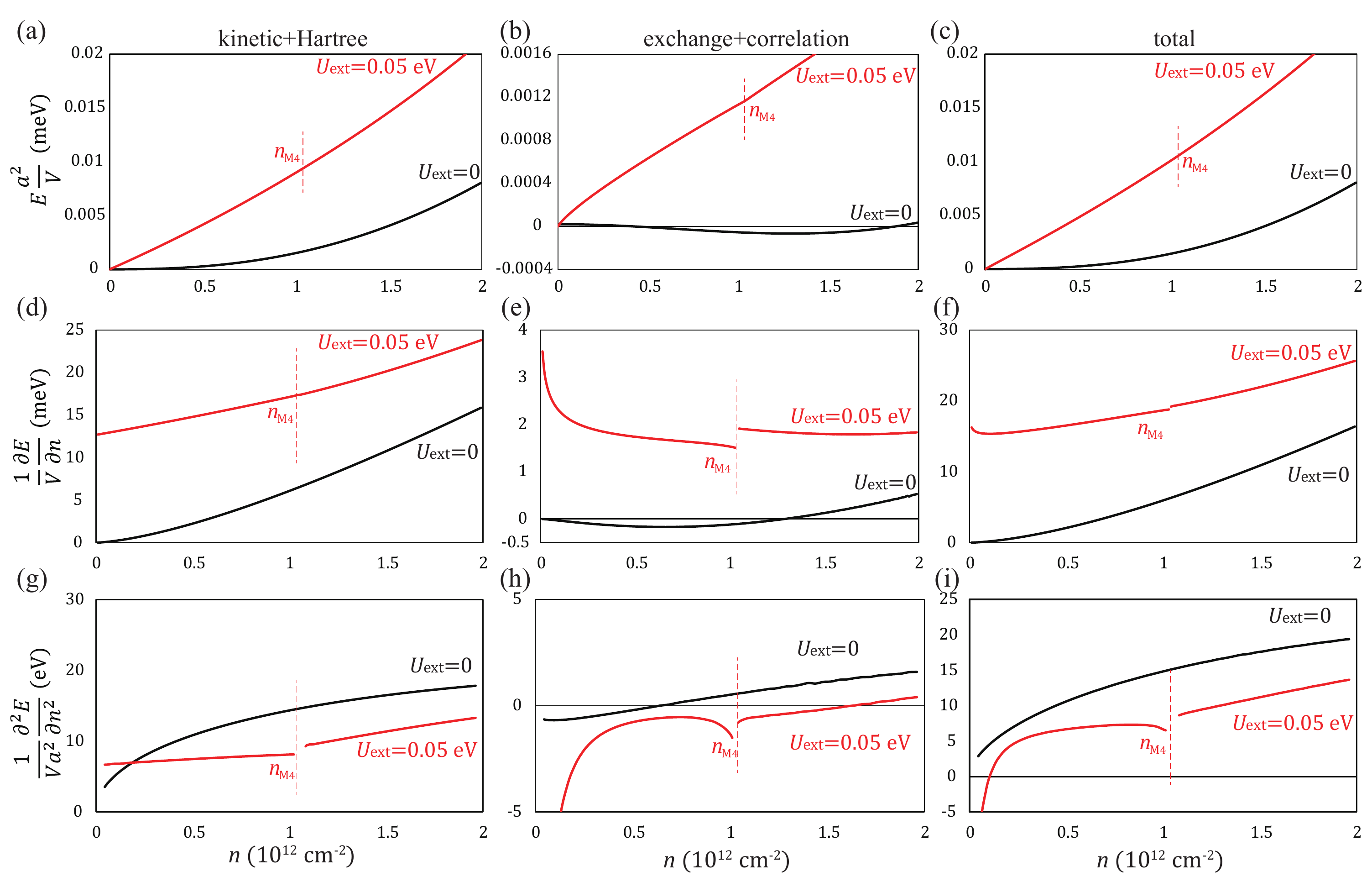}
\caption{
Density dependence of (a) the kinetic plus Hartree energy, (b) the exchange plus correlation energy, (c) the total energy in ABC trilayer graphene for $U_{\rm ext}=0$ (black) and $U_{\rm ext}=0.05$ eV (red), (d), (e), and (f) their first derivatives, and (g), (h), and (i) their second derivatives with respect to the density. 
Note that the second derivatives are related to the electronic compressibility as $\kappa^{-1}={n^2\over V}{\partial^2 E_{\rm tot}\over \partial n^2}$. 
The red vertical dashed lines represent the density $n_{\rm M_4}\approx 1.04\times 10^{12}$ cm$^{-2}$ that fills the Mexican hat structure of the ${\rm P}_4$ state when $U_{\rm ext}=0.05$ eV.
For finite electric fields, the exchange-correlation energies show clearly negative compressibilities up to densities above $n_{\rm M_4}$, while the instability from total energies is expected at low densities near the band edges.
Here, we use the effective fine structure constant $\alpha={e^2\over \epsilon \hbar v}=1$, where $\epsilon$ is the effective dielectric constant.
} 
\label{fig:fig2}
\end{figure*}

While previous studies of flavor polarization have been carried out at the level of Hartree-Fock approximation~\cite{Huang2022,Xie2023}, in this Research Letter we calculate the exchange-correlation energy including the Coulomb correlations within the RPA to obtain the corresponding phase diagram of the spontaneous spin-valley polarization.
Our findings reveal that the exchange interaction between chiral electrons is the driving force behind the emergence of spin-valley polarized states, while the Mexican hat band structure arising from the perpendicular electric field and the correlation effects jointly determine the transition point.
The ground state is determined by comparing the total energies corresponding to phases ${\rm P}_1 \textendash {\rm P}_4$, and we will later discuss different possible intermediate mixed states with unequal flavor concentrations in the Discussion section.
The general tendency towards (against) spontaneous flavor polarization can be predicted from the density dependence of its energy as concave down (up)
as depicted in Fig.~\ref{fig:fig1}(b)
where the system will tend towards (against) spontaneous spin-valley flavor polarization. 
The electronic compressibility is therefore closely connected to spontaneous flavor polarization
and can serve as a powerful thermodynamic probe of the electron-electron interaction effect. 
The inverse electronic compressibility can be expressed as $\kappa^{-1}=n^2 {\partial\mu / \partial n}$, where $n$ is the carrier density, 
$\mu={\partial E_{\rm tot} / \partial N}$ is the chemical potential, $E_{\rm tot}$ is the total ground-state energy, and $N$ is the total number of particles in the system.

\begin{figure*}[tb]
\includegraphics[width=0.95\linewidth]{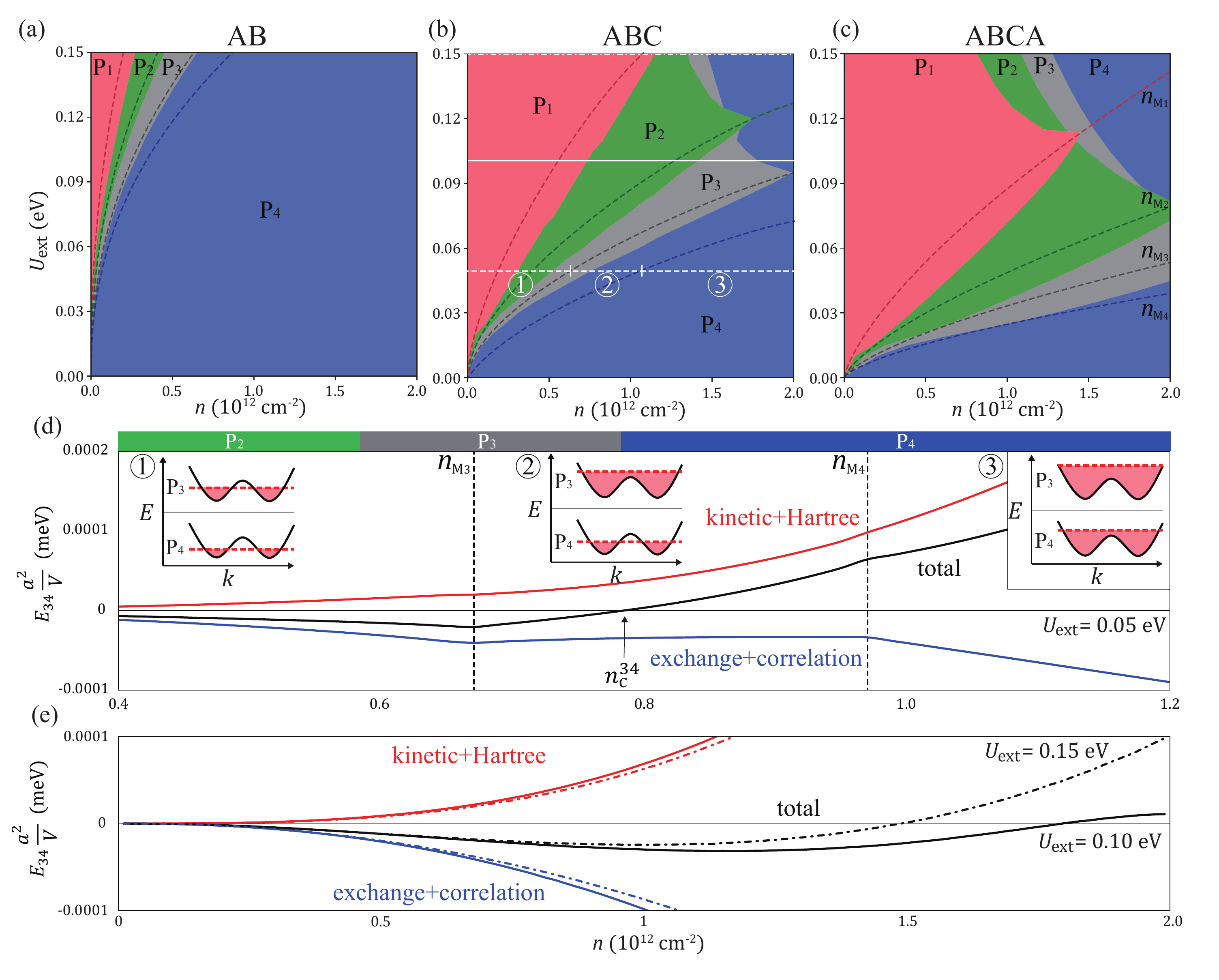}
\caption{
Phase diagrams of (a) AB, (b) ABC, and (c) ABCA stacked multilayer graphene. (d) and (e) Energy difference between the ${\rm P}_3$ and ${\rm P}_4$ states for ABC trilayer graphene for (d) $U_{\rm ext}=0.05$ eV following the white dashed line in (b), and (e) $U_{\rm ext}=0.1$ eV (solid lines) and $U_{\rm ext}=0.15$ eV (dash-dotted lines) following the white solid and dash-dotted lines in (b), respectively. The insets in (d) show the low-energy conduction band and the Fermi energy of the occupied flavors for the ${\rm P}_3$ and ${\rm P}_4$ states 
in regions \raisebox{.5pt}{\textcircled{\raisebox{-.9pt} {1}}}, \raisebox{.5pt}{\textcircled{\raisebox{-.9pt} {2}}}, and \raisebox{.5pt}{\textcircled{\raisebox{-.9pt} {3}}} in (b). In (d), these regions are separated by the vertical dashed lines at $n=n_{\rm M_3}$ and $n=n_{\rm M_4}$ that fill the Mexican hat structure of the ${\rm P}_3$ and ${\rm P}_4$ states, respectively, when $U_{\rm ext}=0.05$ eV.
Here, we use $\alpha=1$.
} 
\label{fig:fig3}
\end{figure*}

{\em Density dependence of the energies and their derivatives.}
To calculate the ground-state energy, we consider the contributions from the kinetic, Hartree, exchange, and correlation energies. In the presence of a perpendicular electric field $E_{\rm ext}$, we start by obtaining a mean-field band structure that takes into account the effect of the self-consistent Hartree potential, as well as the energy gap opening due to $E_{\rm ext}$. Next, we compute the exchange-correlation energy by using the integration-over-coupling-constant method within the RPA. Further details regarding the derivation can be found in the Supplemental Material \cite{SM}.
Figure \ref{fig:fig2} shows the density dependence of the kinetic plus Hartree energy, the exchange plus correlation energy, the total energy, and their first and second derivatives with respect to the density in ABC-stacked trilayer graphene for $U_{\rm ext}=0$ and $U_{\rm ext}=0.05$ eV, respectively, where $U_{\rm ext}=eE_{\rm ext}d$ with the interlayer separation $d=3.35$ $\rm \AA$.

First, let us consider the $U_{\rm ext}=0$ case. In our calculations, we used the full bands continuum Hamiltonian, but here our discussion is based on 
the low-energy C2DES model of rhombohedral multilayer graphene with chirality index $J$ which coincides with layer number \cite{min2008a,min2008b}:
\begin{equation}
\label{eq:Hamiltonian_C2DES}
{\cal H}_J(\bm{k})=t_{\perp} \left(
\begin{array}{cc}
0 & \left({\hbar v k_{-}\over t_{\perp}}\right)^J \\
\left({\hbar v k_{+}\over t_{\perp}}\right)^J & 0 \\
\end{array}
\right),
\end{equation}
where $k_{\pm}=k_x\pm i k_y$, $v$ is the in-plane Fermi velocity of monolayer graphene, and $t_{\perp}$ is the nearest-neighbor interlayer hopping. The eigenenergies of Eq.~(\ref{eq:Hamiltonian_C2DES}) are given by $\varepsilon_{s,\bm{k}}=s t_{\perp} \left(\hbar v |\bm{k}| / t_{\perp}\right)^J$ with $s=\pm 1$ for positive and negative energy states, respectively.
For $U_{\rm ext}=0$, the Hartree energy is zero, and the kinetic energy per particle is given by ${E_{\rm kin}\over N}={2\over J+2}\varepsilon_{\rm F}$, where $\varepsilon_{\rm F}$ is the Fermi energy. Since $\varepsilon_{\rm F}\sim k_{\rm F}^J$ and $n\sim k_{\rm F}^2$, where $k_{\rm F}$ is the Fermi wave vector, we have $E_{\rm kin}\sim k_{\rm F}^{J+2}\sim n^{J+2\over 2}$, which is concave up with respect to the density $n$. Therefore the kinetic energy does not favor spin-valley flavor polarization.
In contrast, the exchange energy per particle is given by ${E_{\rm ex}\over N}=C_1{e^2 \over \epsilon_0}k_{\rm F}$, where $C_1$ is a coefficient with weak density dependence such that $C_1<0$ for $J\ge 2$ ($C_1>0$ for $J=1$) due to the dominant intraband (interband) exchange interaction \cite{Jang2015}. This means that for $J\ge 2$, $E_{\rm ex}\sim -k_{\rm F}^3\sim -n^{3\over 2}$, which is concave down with respect to $n$, and thus the exchange energy favors flavor polarization.
From the power-law dependence, we find that at small carrier densities, the exchange energy is dominant over the kinetic energy, and there is a tendency toward flavor polarization. However, at sufficiently large densities, the kinetic energy becomes dominant over the exchange energy, and there is a tendency toward the normal phase.
When correlation effects are included, the tendency toward flavor polarization is reduced due to the concave-up dependence of the correlation energy on density. The exchange contribution to the electronic compressibility is negative, but when all contributions are combined, the electronic compressibility remains positive. Thus ABC trilayer graphene remains in the normal phase for $U_{\rm ext}=0$, as indicated by the black line in Fig.~\ref{fig:fig2}(i).

This scenario changes when $U_{\rm ext}\neq 0$, where the system develops a characteristic ``Mexican hat" structure that we obtain self-consistently by considering the kinetic and Hartree contributions. As the density $n$ increases, the Fermi energy also increases, causing the Fermi surface to evolve from a disk with a concentric hole to a fully filled disk. In our approximation, the Fermi energy corresponds to the first derivative of the kinetic plus Hartree energy with respect to $n$; so the second derivative of the energy remains positive even in the presence of $U_{\rm ext}$, indicating a tendency toward the normal phase.
The exchange energy is also influenced by a finite $U_{\rm ext}$. A perpendicular electric field causes some pseudospins in the conduction and valence bands to align oppositely along the $z$ direction. This alignment results in an increase of the interband contribution to the exchange energy with increasing density, up to a certain density where the oppositely aligned pseudospins induce a maximum exchange energy. Beyond this density, the intraband contribution to the exchange energy dominates, causing the exchange contribution to decrease with increasing density, similar to the $U_{\rm ext}=0$ case. Additionally, as the pseudospins in the conduction band become more aligned along the same direction due to the external field, the second derivative of the exchange energy becomes more negative, making the exchange energy more concave down and enhancing the tendency toward spin-valley polarization.
When the density $n$ crosses the density $n_{\rm M}$ that fills the Mexican hat structure, the negative intraband exchange contribution from electrons near $\bm{k}=0$ becomes absent, resulting in a less negative contribution to the exchange energy. This causes a jump in the first derivative of the exchange energy, as shown in Fig.~\ref{fig:fig2}(e). We note that typically the correlation reduces the exchange effects.
At low densities, the electronic compressibility $\kappa$ of the system becomes negative for $U_{\rm ext}\neq 0$, which means that the system no longer remains in the normal phase and instead enters a flavor polarized state. However, at high enough densities, the kinetic energy associated with large $\bm{k}$ becomes dominant, transitioning back to the normal phase.

{\em Phase diagram of multilayer graphene.}
The density-dependent total energy analysis shows an inherent tendency of the exchange interaction towards spin-valley flavor polarization.
In the following we present the flavor polarization phase diagram of bilayer, trilayer, and tetralayer rhombohedral multilayer graphene as a function of carrier density $n$ and external potential $U_{\rm ext}$, where the total energies include the RPA correlations. To calculate the exchange-correlation energy in multilayer graphene, we use the rotational transformation of the chiral wave function to obtain the chiral wave function at any angle from the wave function obtained at a given angle, greatly facilitating the calculations~\cite{Jang2015}. 
Figures \ref{fig:fig3}(a)--3(c) show the phase diagrams of AB, ABC, and ABCA stacked multilayer graphene as a function of $n$ and $U_{\rm ext}$. To further understand these phase diagrams, let us examine the band structure change of ABC trilayer graphene near the phase boundaries along the white dashed line in Fig.~\ref{fig:fig3}(b) when we use a fixed $U_{\rm ext}=0.05$~eV.
For example, the band fillings for the ${\rm P}_3$ and ${\rm P}_4$ states near the phase boundaries have qualitatively different characters. For a given equal density in the ${\rm P}_3$ state, the Mexican hat band structure is fully filled beyond the top of the hat near the Dirac points $K^{(\prime)}$, 
while for the ${\rm P}_4$ state the electrons fill up to lower energies.
Because in the ${\rm P}_3$ phase the (negative intraband) exchange contribution from electrons near $\bm{k}=0$ is already filled, an increase in the carrier density results in a less negative contribution to the exchange energy compared with the ${\rm P}_4$ phase, where the electrons are mainly filling the ring-shaped band edges and the exchange energy gain is greater [see the insets in Fig.~\ref{fig:fig3}(d)].
This implies that in practice there will be a flavor polarization transition just after the Mexican hat structure is completely filled because the exchange-driven flavor polarization is partly countered by the Coulomb-correlation-driven screening.
The energy difference $E_{34}$ between the ${\rm P}_3$ and ${\rm P}_4$ states becomes positive after the critical density $n_{\rm c}^{34}$ as indicated by the black arrow in Fig.~\ref{fig:fig3}(d). As the perpendicular field is increased to $U_{\rm ext} = 0.1$~eV [see Fig.~\ref{fig:fig3}(e)], the density required to fill the Mexican hat structure ($n_{\rm M_3}$) and the phase transition critical density ($n_{\rm c}^{34}$) also increase.
In the higher-density region, the difference between the two densities $n_{\rm M_3}$ and $n_{\rm c}^{34}$ becomes smaller since the kinetic plus Hartree energy is dominant over the exchange-correlation energy. Thus $n_{\rm M_3}$ and $n_{\rm c}^{34}$ eventually merge at the same point, showing a kink structure in the phase diagram. For a sufficiently high $U_{\rm ext}$ above the kink point, the Mexican hat structures of both the ${\rm P}_3$ state and the ${\rm P}_4$ state are partially filled, and $E_{34}$ increases by $U_{\rm ext}$ as shown in Fig.~\ref{fig:fig3}(e), resulting in a decreased $n_{\rm c}^{34}$. 
The other phase boundaries in the phase diagram can be discussed using similar arguments.

{\em Discussion.}
The calculated phase diagram in Fig.~\ref{fig:fig3} captures the main features of the carrier density and electric field dependence of the spontaneous spin-valley polarization recently observed in rhombohedral trilayer graphene \cite{Zhou2021a} and bilayer graphene~\cite{Zhou2022}. 
The explicit inclusion of the RPA correlations allows us to achieve agreement with experiments with a conventional value of the dielectric constant of $\epsilon_r \approx 2.6$ corresponding to the effective interaction strength $\alpha = 1$, rather than using unrealistic screening constants to compensate the overestimation of the exchange interaction. Instead, in our calculation
the Coulomb correlations cause the phase boundaries to appear close to the densities where the top of the Mexican hat is filled completely for a given flavor polarization.
Our theory is expected to be valid in the limit of strong electric fields and large densities, showing that the phase boundary slopes agree closely with experiments in this regime.

So far, our analysis has been based on pure flavor polarized states rather than intermediate mixed states with unequal flavor concentrations. Further discussion of the mixed states can be found in the Supplemental Material~\cite{SM}.
%
%
%
Below, we discuss some details left out in our study.
Firstly, our work assumes an isotropic circular symmetry around the Dirac points neglecting the trigonal warping of the bands to reduce the computational load. 
These effects are expected to be small away from $n=0$ and $U_{\rm ext}=0$, thus allowing us to capture the spontaneous spin-valley polarized phases that are manifested away from this limit.
Secondly, our current calculation does not include any anisotropy term that favors a certain flavor polarization over another. 
The absence of a flavor anisotropy term may also explain the appearance of the ${\rm P}_3$ state in our current calculations, which is absent in experiments. 
We leave the analysis of flavor degeneracy breaking as an open question for future research.

In summary, we have discussed the spin-valley flavor polarization in bi-, tri-, and tetralayer rhombohedral graphene in light of the long-range Coulomb correlations neglected in the literature. We have shown that while the exchange interaction together with the chiral bands drives the instability towards flavor polarization, the screening due to Coulomb correlations 
brings the phase boundary points to lie close to the carrier densities that fill the top of the Mexican hat band structure. 
The carrier density, external electric field, and band chirality 
influence the phase diagram of the spin-valley flavor polarized states in rhombohedral multilayer graphene, suggesting a greater tendency of flavor polarization in systems with larger band chirality and, therefore, a larger number of layers.


\acknowledgments
This work was supported by the National Research Foundation of Korea (NRF) grants funded by the Korea government (MSIT) (Grants No. 2018R1A2B6007837 and No. 2023R1A2C1005996) and the Creative-Pioneering Researchers Program through Seoul National University (SNU).  Y.P. acknowledges support from Samsung Science and Technology Foundation Grant No. SSTF-BA1802-06.
J.J. acknowledges support from NRF (Grant No. 2020R1A2C3009142), KISTI (Grant No. KSC2022-CRE-0514), the resources of Urban Big data and AI Institute (UBAI) at UOS, and the Korean Ministry of Land, Infrastructure and Transport (MOLIT) from the Innovative Talent Education Program for Smart Cities.

\clearpage
\widetext
\setcounter{section}{0}
\setcounter{equation}{0}
\setcounter{figure}{0}
\setcounter{table}{0}
\renewcommand{\theequation}{S\arabic{equation}}
\renewcommand\thefigure{S\arabic{figure}} 
\setcounter{page}{1}
\begin{center}
\textbf{\large Supplemental Material for ``Chirality and correlations in the spontaneous spin-valley polarization of rhombohedral multilayer graphene"}
\end{center}
\begin{center}
{Yunsu Jang,$^{1,2}$ Youngju Park,$^{3}$ Jeil Jung,$^{3,4,\ast}$ and Hongki Min$^{1,2\dagger}$}\\
\emph{$^{1}$ Department of Physics and Astronomy, Seoul National University, Seoul 08826, Korea}\\
\emph{$^{2}$ Center for Theoretical Physics, Seoul National University, Seoul 08826, Korea}\\
\emph{$^{3}$ Department of Physics, University of Seoul, Seoul 02504, Korea}\\
\emph{$^{4}$ Department of Smart Cities, University of Seoul, Seoul 02504, Korea}
\end{center}

\section{Calculation of the exchange-correlation energy using a self-consistent Hartree band structure}
\label{sec:exchange-correlation_energy}

When a perpendicular electric field is applied, it causes a redistribution of the charge density within each layer. This redistribution affects the electric fields between the layers. To take this effect into account, we calculate the Hartree potential energy self-consistently. This process is equivalent to solving the classical Poisson equation in electrostatics~ \cite{SM_jang2019optical}. From the resulting band structure and wave function, we calculate the exchange-correlation energy using the integration-over-coupling-constant method within the random phase approximation (RPA).


\begin{figure*}[ht!]
\includegraphics[width=0.5\linewidth]{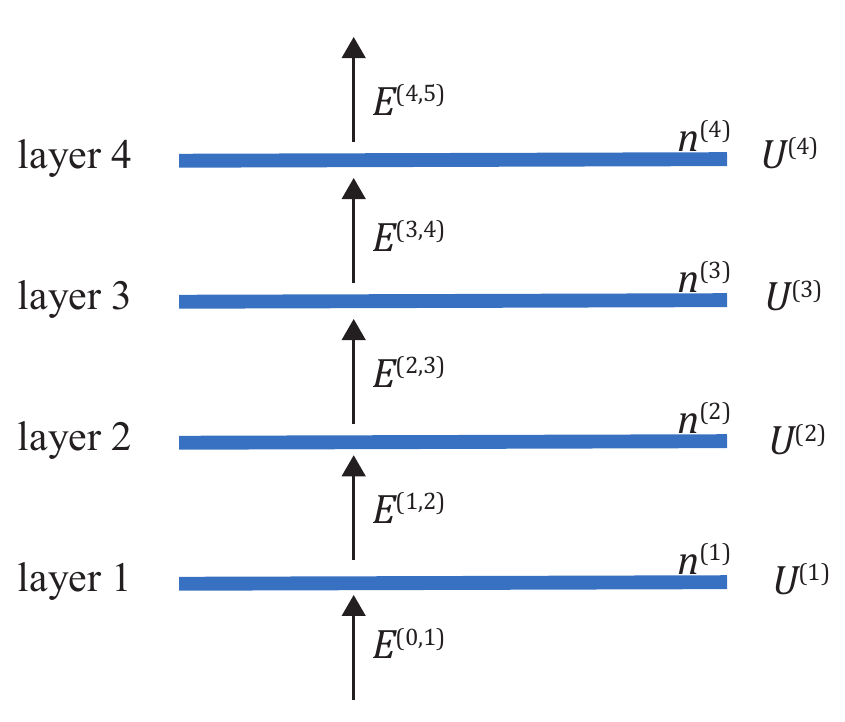}
\caption{
Schematic illustration of tetralayer graphene in the presence of a perpendicular electric field. Here, $n^{(i)}$ and $U^{(i)}$ are the charge density and the potential energy in the $i$th layer, and $E^{(i-1,i)}$ is the electric field between the $i$th and ($i$+1)th layers along the $z$ direction.
} 
\label{fig:sfig1}
\end{figure*}

From Gauss's law, the electric field $E^{(i,i+1)}$ between the $i$th and ($i$+1)th layers along the $z$ direction is given by (see Fig.~\ref{fig:sfig1})
\begin{equation}
E^{(i,i+1)}-E^{(i-1,i)}= 4\pi(-e) n^{(i)}/\epsilon,
\end{equation}
where $n^{(i)}$ is the charge density in the $i$th layer and $\epsilon$ is the effective dielectric constant. The top-gate density $n^{\rm top}$ and bottom-gate density $n^{\rm bottom}$ (not shown in Fig.~\ref{fig:sfig1}) determine the electric field above the top layer as $E^{\rm top}=4\pi e n^{\rm top}/\epsilon$ and that below the bottom layer as $E^{\rm bottom}=-4\pi e n^{\rm bottom}/\epsilon$, respectively. 
Because the whole system including the top and bottom gates is charge neutral, the sum of the top-gate, bottom-gate and layer charge densities must be zero:
\begin{equation}
n^{\rm top}+n^{\rm bottom}+\sum_i n^{(i)}=0.
\end{equation}
For a given external electric field $E_{\rm ext}={1\over 2}\left(E^{\rm top}+E^{\rm bottom}\right)$, we can obtain the electric field contribution due to induced charge densities as
\begin{equation}
E_{\rm ind}^{(i-1,i)} = E^{(i-1,i)}-E_{\rm ext}.
\end{equation}

To account for the effect of the perpendicular electric field on the charge density in each layer, we solve the Poisson equation self-consistently to obtain the Hartree potential energy. The potential energy $U^{(i)}$ in the $i$th layer can be written as

\begin{equation}
\label{eq:layer_potential_energy}
U^{(i)} - U^{(i-1)} = e E^{(i,i-1)} d,
\end{equation}
where $d=3.35$ $\rm \AA$ is the interlayer separation between graphene layers.
Similarly, we define $U_{\rm ind}^{(i)}$ and $U_{\rm ext}^{(i)}$ by replacing $E^{(i,i-1)}$ in Eq.~(\ref{eq:layer_potential_energy}). 
Here we set the average of the band energy to zero at large momentum ($|\bm{k}|\rightarrow \infty$), which corresponds to $\sum_i U^{(i)}=0$.
We use an effective dielectric constant of $\epsilon\approx 2.6$, or equivalently, $\alpha={e^2\over \epsilon \hbar v}\approx 1$, to account for the effect of substrates.
 
For given $U_{\rm ind}^{(i)}$ and $U_{\rm ext}^{(i)}$, we can construct a mean-field Hamiltonian including the effect of the Hartree potential as
\begin{equation}
\label{eq:mean_field_Hamiltonian}
H_{\rm MF}=H_{\rm kin}^{(0)}+H_{\rm ext}+H_{\rm ind},
\end{equation}
where $H_{\rm kin}^{(0)}$ is the band Hamiltonian of rhombohedral multilayer graphene, $H_{\rm ext}={\rm {diag}}(U_{\rm ext}^{(1)}\mathbb{I}_{2}, U_{\rm ext}^{(2)}\mathbb{I}_{2}, ..., U_{\rm ext}^{(N)}\mathbb{I}_{2})$ is the Hamiltonian for the external potential energy and $H_{\rm ind}={\rm {diag}}(U_{\rm ind}^{(1)}\mathbb{I}_{2}, U_{\rm ind}^{(2)}\mathbb{I}_{2}, ..., U_{\rm ind}^{(N)}\mathbb{I}_{2})$ for the induced potential energy which corresponds to the Hartree contribution. By solving the mean-field Hamiltonian in Eq.~(\ref{eq:mean_field_Hamiltonian}), we can obtain the eigenenergies $\varepsilon_{s,\bm{k}}$ and the corresponding wave functions $\psi_{s,\bm{k}}$ for the band index $s$ and wave vector $\bm{k}$.

The kinetic energy that takes into account the external potential is given by
\begin{equation}
\label{eq:kinetic_energy}
E_{\rm kin} = g_{\rm sv} \sum_{s} \int \frac{d^{2}k}{(2\pi)^2}  \left<\psi_{s} \right| H_{\rm kin}^{(0)}+H_{\rm ext} \left| \psi_{s}  \right> f_{s,\bm{k}},
\end{equation}
where $f_{s,\bm{k}}$ is the Fermi distribution function for the band $s$ and wave vector ${\bm k}$ and $g_{\rm sv}=g_{\rm s}g_{\rm v}=4$ is the spin-valley degeneracy.

The Hartree energy due to the induced potential is given by
\begin{equation}
\label{eq:Hartree_energy}
E_{\rm Hartree} = \frac{g_{\rm sv}}{2} \sum_{s} \int \frac{d^{2}k}{(2\pi)^2} \left<\psi_{s} |H_{\rm ind}| \psi_{s} \right> f_{s,\bm{k}}.
\end{equation}
Note that Eq.~(\ref{eq:Hartree_energy}) includes the factor 1/2 to eliminate double counting.




Starting from the mean-field band structure obtained from the self-consistent Hartree approximation described above, we can obtain the exchange and the RPA correlation energies using the integration-over-coupling constant method assuming the thin-film limit as \cite{SM_Giuliani2005,SM_Jang2015,SM_Polini2007}
\begin{eqnarray}
\frac{E_{\rm ex}}{V}&=&-{\hbar \over 2} \int {d^2q \over (2\pi)^2} \int_0^{\infty} {d\omega \over \pi} V_{\bm{q}}\delta \Pi_0(\bm{q},i\omega), \label{eq:exchange_correlation_energy1} \\
\frac{E_{\rm corr}}{V}&=&{\hbar \over 2} \int {d^2q \over (2\pi)^2} \int_0^{\infty} {d\omega \over \pi} \bigg[ V_{\bm{q}}\delta \Pi_0(\bm{q},i\omega) +\ln\left| {1-V_{\bm{q}}\Pi_0(\bm{q},i\omega) \over 1-V_{\bm{q}}\left.\Pi_0(\bm{q},i\omega)\right|_{n=0}} \right| \bigg],
\label{eq:exchange_correlation_energy2}
\end{eqnarray}
where $V_{\bm{q}}=2\pi e^2/(\epsilon q)$ is the two-dimensional Coulomb interaction, $\Pi_{\rm 0}$ is the noninteracting electron density-density response function for imaginary frequency defined by
\begin{equation}
\Pi_{0}(\bm{q},i\omega)=g_{\rm sv}  \sum_{s,s'}\int {d^2 k \over (2\pi)^2} {f_{s,\bm{k}}-f_{s',\bm{k+q}} \over  i\hbar\omega + \varepsilon_{s,\bm{k}}-\varepsilon_{s',\bm{k+q}}} F_{\bm{k},\bm{k+q}}^{s,s'},
\end{equation}
$\delta\Pi_0(\bm{q},i\omega) = \Pi_0(\bm{q},i\omega) - \left. \Pi_0(\bm{q},i\omega) \right|_{n=0}$,
and $F^{s,s'}_{\bm{k},\bm{k+q}} = \left| \left< \psi_{\bm{k},s} |\psi_{\bm{k'},s'} \right> \right|^{2}$ is the wavefunction overlap factor.

In spin-valley polarized states, electrons no longer occupy the four spin-valley flavors equally but tend to occupy some of the four flavors, thus each spin-valley flavor needs to be considered separately as
\begin{eqnarray}
     g_{\rm sv} & \rightarrow & \sum_{ \xi}, \\
     f_{s,\bm{k}} & \rightarrow & f_{s,\bm{k}}^{\xi},
\end{eqnarray}
where ${ \xi}$ is a spin-valley index. In the case of the ${\rm P}_3$ state, for example, the density is equally distributed for $\xi=1,2,3$ while zero for ${\xi}=4$. Accordingly, the kinetic and Hartree energies in Eqs.~(\ref{eq:kinetic_energy}) and (\ref{eq:Hartree_energy}) are modified respectively as
\begin{eqnarray}
    E_{\rm kin}&\rightarrow&\sum_{\xi} \sum_{s} \int \frac{d^{2}k}{(2\pi)^2}  \left<\psi_{s} \right| H_{\rm kin}+H_{\rm ext} \left| \psi_{s}  \right> f_{s,\bm{k}}^{\xi},\\
    E_{\rm Hartree}&\rightarrow& \frac{1}{2} \sum_{\xi}  \sum_{s} \int \frac{d^{2}k}{(2\pi)^2} \left<\psi_{s} |H_{\rm ind}| \psi_{s} \right> f_{s,\bm{k}}^{\xi}.
\end{eqnarray}

On the other hand, the form of the exchange and correlation energies in Eqs. (\ref{eq:exchange_correlation_energy1}) and (\ref{eq:exchange_correlation_energy2}) remain the same but $\Pi_0$ is modified as
\begin{equation}
\Pi_{0}(\bm{q},i\omega) \rightarrow \sum_{\xi}  \sum_{s,s'}\int {d^2 k \over (2\pi)^2} {f_{s,\bm{k}}^{\xi}-f_{s',\bm{k+q}}^{\xi} \over  i\hbar\omega + \varepsilon_{s,\bm{k}}-\varepsilon_{s',\bm{k+q}}} F_{\bm{k},\bm{k+q}}^{s,s'}.
\end{equation}
In this paper, we set the energy at $n=0$ and $U_{\rm ext}=0$ as the zero of energy.

\section{Intermediate mixed states in the phase diagram}
\label{sec:intermediate_mixed_states}


In the experiments by Zhou {\it et al.}~\cite{SM_Zhou2021a,SM_Zhou2022}, there were intermediate mixed states between the pure flavor polarized phases that were not captured by our calculated phase diagram in Fig.~\ref{fig:fig3}. This is because we determined the phase diagram by comparing the total energy of each phase assuming for simplicity that the occupied flavors have equal concentrations. 
However, an intermediate mixed state with unequal flavor concentrations is also possible. For example, between the ${\rm P}_1$ and ${\rm P}_2$ states, a state in which two flavors are occupied with unequal concentrations is possible if it has a lower energy compared to those of pure ${\rm P}_1$ and ${\rm P}_2$.

\begin{figure*}[htb]
\includegraphics[width=0.75\linewidth]{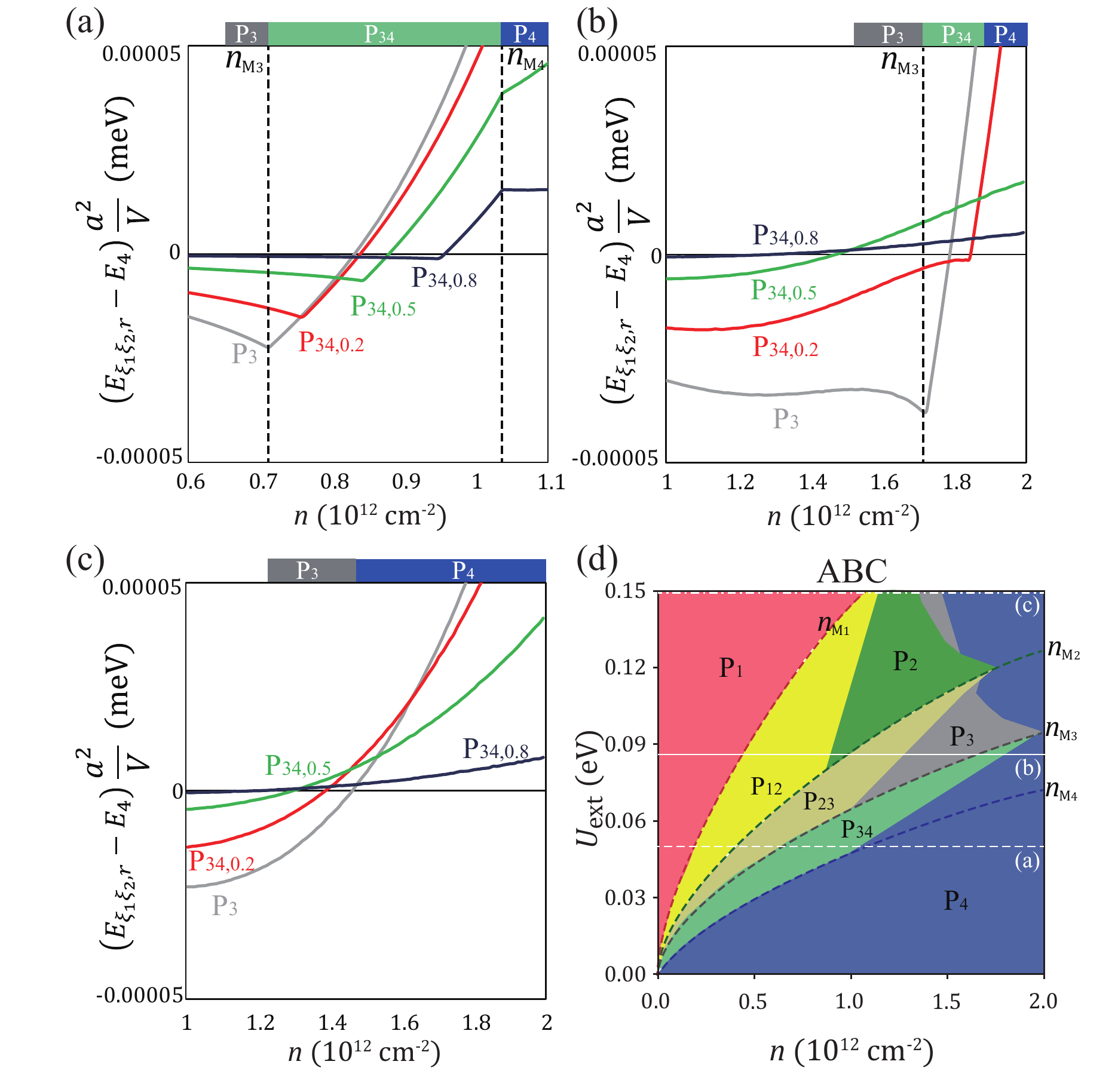}
\caption{Total energy difference between the intermediate mixed state ${\rm P}_{34,r}$ and the pure state ${\rm P}_4$ for (a) $U_{\rm ext}=0.05$ eV, (b) $U_{\rm ext}=0.085$ eV and (c) $U_{\rm ext}=0.15$ eV with $r=0.2, 0.5, 0.8$. (d) Schematic illustration of the pure and intermediate mixed states in the phase diagram of ABC trilayer graphene.
} 
\label{fig:sfig2}
\end{figure*}

In the intermediate mixed state, the occupied flavors can have different densities from one another. We denote the intermediate mixed state between the two pure states $P_{{\xi}_1}$ and $P_{{\xi}_2}$ with the mixing ratio $r$ as $P_{{\xi}_1{\xi}_2,r}$ where we assume  ${\xi}_1, {\xi}_2 = 1,2,3,4$ with ${\xi}_1 < {\xi}_2$. We divide the spin-valley flavors into three groups with densities $n_1$, $n_2$ and $n_3=0$, and degeneracies ${\xi}_1$, ${\xi}_2 -{\xi}_1$ and $4-{\xi}_2$, respectively. Then the densities $n_1$ and $n_2$ are given by $n_1=(1-r)(n/{\xi}_1) + r(n/{\xi}_2)$ and $n_2=r(n/{\xi}_2)$, respectively, satisfying ${\xi}_1 n_1 + ( {\xi}_2 -{\xi}_1 ) n_2 = n$. 
When $r=0$ ($r=1$), the intermediate mixed state $P_{{\xi}_1{\xi}_2,r}$ corresponds to the pure state $P_{{\xi}_1}$ ($P_{{\xi}_2}$). 
For example, in the case of $P_{12,0.2}$, the densities of each spin-valley can be expressed as  $\{n^{\xi}\}= (n_1,n_2,0,0)$ where $n_1=0.8(n/p_1) + 0.2(n/p_2) = 0.9n$ and $n_2=0.2(n/p_2)= 0.1n$. 

Figure \ref{fig:sfig2}(a) shows the total energy difference between the intermediate mixed state ${\rm P}_{34,r}$ and the pure state ${\rm P}_4$ for $r=0.2, 0.5, 0.8$ when $U_{\rm ext}=0.05$ eV. 
For an intermediate mixed state ${\rm P}_{34,r}$ with $0<r<1$, there appears a region which has a lower total energy than that of the pure states ${\rm P}_3$ and ${\rm P}_4$ because of the rapid change in the total energy at the density that fills the Mexican hat structure, $n_{{\rm M}_\xi}$. (Here, $r$ can be determined by minimizing the total energy for given $U_{\rm ext}$ and $n$.) This means that in the region $n_{{\rm M}_3}<n<n_{{\rm M}_4}$, the total energy of an intermediate mixed state becomes lower than that of the pure states. Similarly, intermediate mixed states appear in the region $n_{{\rm M}_1}<n<n_{{\rm M}_2}$ and $n_{{\rm M}_2}<n<n_{{\rm M}_3}$, thus pure states can occur only in the regions $n\le n_{{\rm M}_1}$ and $n\ge n_{{\rm M}_4}$ as well as on the lines $n=n_{{\rm M}_2}$ and $n=n_{{\rm M}_3}$ in the phase diagram. 
As $U_{\rm ext}$ increases, however, it is possible that the pure state ${\rm P}_4$ can appear in the region $n_{{\rm M}_3}<n<n_{{\rm M}_4}$ when the minimum of the total energy of ${\rm P}_{34,r}$ is higher than the total energy of ${\rm P}_4$, as shown in Fig.~\ref{fig:sfig2}(b) for $U_{\rm ext}=0.085$ eV. Then pure-state regions and intermediate-state regions can occur side by side in the phase diagram. For large $U_{\rm ext}$ (above the potential at the kink structure in the phase diagram), the density that the minimum energy occurs is no longer accompanied by $n_{{\rm M}_\xi}$ and a pure state always has lower energy than that of an intermediate mixed state, as shown in Fig.~\ref{fig:sfig2}(c) for $U_{\rm ext}=0.15$ eV. Figure \ref{fig:sfig2}(d) shows a schematic illustration of the pure and intermediate mixed states in the phase diagram of ABC trilayer graphene.
In the current calculation, intermediate mixed states tend to appear more dominantly over pure states in the low-field region of the phase diagram compared to the experiment.
This is due to the fact that we use the wave function and Fermi surface obtained from the self-consistent Hartree approximation to calculate the exchange-correlation energy taking into account the effect of the gap opening, which could result in a sudden change in the total energy at $n_{\rm M}$, as indicated by a discontinuity in the slope of the total energy at $n_{\rm M}$ (see Figs.~\ref{fig:fig2}(d) and \ref{fig:fig2}(f) in the main text). These results could be slightly improved by allowing a degree of freedom to change the inner and outer radii of the Fermi surface for a given density, but with a significant increase in computational cost. In order to correctly determine the intermediate mixed states in the phase diagram, we need to include the effect of the exchange-correlation energy in the band structure, which is beyond the scope of the current approximation.

\end{document}